\documentstyle{article}
\def\graphic #1#2#3#4#5{

    \noindent
    \centerline{\hrulefill}
    \leftline{\hbox to#1{\special{anisoscale #3, #1 #2}\hfil}}
    \vspace*{#2} \relax
    \vskip -3.9 cm
    \hskip 4.8 cm
    {\large \bf Universidade do Estado do Rio de Janeiro }
    \newline

    \vskip -0.25 cm
    \hskip 7.5 cm
    {\large \bf Instituto de F{\'\i}sica }

    \vskip 1 cm
    \hskip 7.5 cm
    {\large IF-UERJ-#4 }

    \hskip 7.5 cm
    {\large Preprint}

    \hskip 7.5 cm
    {\large #5 }

    \medskip
    \noindent
    \hrulefill

    \vskip 2.9 cm
    }
\def\({\c c}
\def\|{\'\i}

\def\dgraphic #1#2#3#4{
    \centerline{\hbox to#1{\special{anisoscale #3, #1 #2}\hfil}} % put graph
    \vspace*{#2} \relax         % next text begins after this vspace
    \begin{figure}[h] \caption{#4} \end{figure}  % caption and auto. fig. No.
    }
\def\dtwographic #1#2#3#4{
    \centerline{\hbox to#1{\special{anisoscale #3, #1 #2}\hfil}
                \hbox to#1{\special{anisoscale #4, #1 #2}\hfil}}
    \vspace*{#2} \relax
    }

\def\dthreegraphic #1#2#3#4#5{
    \centerline{\hbox to#1{\special{anisoscale #3, #1 #2}\hfil}
                \hbox to#1{\special{anisoscale #4, #1 #2}\hfil}
                \hbox to#1{\special{anisoscale #5, #1 #2}\hfil}}
    \vspace*{#2} \relax
}

\def\LaTeX{L\kern-.25em\raise.425ex\hbox{a}\kern-.075em\TeX}

\newtheorem{theorem}{Theorem}

\def\ni{\noindent }
\def\eq #1{Eq.(\ref{#1})}
\def\l{\left}
\def\r{\right}
\def\pa{\partial}
\def\fr{\frac}
\def\se #1{sec. \ref{#1}}

\def\nth{$n^{th}$}

\def\y1{\mbox{$y'$}}
\def\yt{\mbox{$y''$}}

\def\yk{\mbox{$y^{(k)}$}}
\def\yn{\mbox{$y^{(n)}$}}
\def\ym{\mbox{$y^{(n-1)}$}}
\def\ymm{\mbox{$y^{(n-2)}$}}
\def\F{{\cal F}}

\def\G{{\cal G}}
\def\H{{\cal H}}

\def\ode{\mbox{\rm ode := \ \ }}

\def\mut{\stackrel{\sim}{\mu}\!}

\def\bigl({\left({\vrule height1.29em width0em depth1.29em}}
\def\bigr){{\vrule height1.29em width0em depth1.29em}\right)}
\def\medl({\left({\vrule height1em width0em depth1em}}
\def\medr){{\vrule height1em width0em depth1em}\right)}
\def\|{\'\i}
\begin{document}
\hspace\parindent
\thispagestyle{empty}

\graphic{2 in}{1.6 in}{uerj.wmf}{2/97}{November 1997}
\centerline{\LARGE \bf {Integrating Factors and ODE Patterns}}

\bigskip
\bigskip
% Authors:
\centerline{\large
E.S. Cheb-Terrab\footnote{Department of Computer Science, Faculty of
Mathematics, University of Waterloo, Ontario, Canada}$^,$\footnote{Symbolic
Computation Group, Departamento de F\|sica
Te\'orica, IF-UERJ.
\newline
\hspace*{.55cm}Available as http://dft.if.uerj.br/preprint/e7-2.tex;
also as http://lie.uwaterloo.ca/odetools/e7-2.tex},
{A.D. Roche}\footnotemark[1] }

\bigskip
\bigskip
\bigskip
\begin{large}
\begin{abstract}

A systematic algorithm for building integrating factors of the form
$\mu(x,\y1)$ or $\mu(y,\y1)$ for non-linear second order ODEs is presented.
When such an integrating factor exists, the algorithm determines it {\it
without solving any differential equations}. Examples of ODEs not having
point symmetries are shown to be solvable using this algorithm. The scheme
was implemented in Maple, in the framework of the {\it ODEtools} package and
its ODE-solver. A comparison between this implementation and other computer
algebra ODE-solvers in tackling non-linear examples from Kamke's book is
shown.

\end{abstract}
\bigskip
\centerline{ \underline{\hspace{6.5 cm}} }

\medskip
\centerline{ {\bf
(Submitted for publication in Journal of Symbolic Computation)} }

\end{large}
\newpage

\section{Introduction}

% Symbolic methods for finding closed form solutions for ordinary
% differential equations (ODEs) have been under development since the advent
% of differential calculus.

>From a practical point of view, when developing solving methods for ODEs,
what one actually does is attempt to determine families of ODEs which can be
transformed into algebraic problems or into simple ODEs such
as\footnote{Throughout this article, we use the notation $y=y(x), \y1=
\frac{d y}{d x}, \yn = \frac {d^n y}{d {x}^n}$.} $\y1 =F(x)$ or $\y1 = F(y)$
by changes of variables or equivalent processes. For high order ODEs, one
hopes that such a simplification of the problem will be possible after
successive reductions of order. Some more powerful schemes are also able to
exploit other information, as for instance integrating factors or the ODE's
symmetries, and so to try a multiple reduction of order at once (see for
instance \cite{maccallum} and \cite{stephani}).

In the specific case of integrating factors, although in principle it can
always be determined whether a given ODE is exact (a total derivative),
there is no known universal scheme for making ODEs exact. Actually, for \nth
order ODEs - as in the case of symmetries - integrating factors are
determined as solutions of an \nth order linear PDE in $n$+1 variables, and
to solve this {\it determining} PDE is a major problem in itself.

Bearing this in mind, this paper presents a method for obtaining integrating
factors of the form $\mu(x,\y1)$ and $\mu(y,\y1)$ for non-linear second
order explicit ODEs\footnote{We say that a second order ODE is in explicit
form when it appears as $\yt-\Phi(x,y,\y1)=0$.},
using a different approach, based only on a computerized analysis of the
pattern of the given ODE. That is, for a given ODE, if an integrating factor
with such a functional dependence exists, the scheme returns the integrating
factor itself without solving any differential equations.

The exposition is organized as follows. In \se{introduction}, the use of
integrating factors for solving ODEs is briefly reviewed. In \se{intfactor},
the scheme for obtaining the aforementioned integrating factors $\mu(x,\y1)$
or $\mu(y,\y1)$ is presented and some examples are given. In \se{mu_and_X},
some aspects of the integrating factor and symmetry approaches are reviewed,
and their complementariness is illustrated with two ODE families not having
point symmetries. Sec. \ref{tests} contains some statistics concerning the
new solving method and the second order non-linear ODEs found in Kamke's
book, as well as a comparison of performances of computer algebra packages
in solving a related subset of these ODEs. In \se{commands} the computer
algebra implementation of the scheme in the framework of the ODEtools
package \cite{odetools2} is outlined, and a description of the package's new
command, {\bf redode}, is presented. Finally, the conclusions contain some
general remarks about this work and its possible extensions.

Aside from this, in the Appendix, a table containing extra information
concerning integrating factors for some of Kamke's ODEs is presented.

\section{Integrating factors and reductions of order}
\label{introduction}

\subsection{First order ODEs}

The idea of looking for an integrating factor ($\mu$) is usually presented
in the framework of solving a given first order ODE, say,

\begin{equation}
\y1=\Phi(x,y)
\label{o1}
\end{equation}

\ni If by multiplying \eq{o1} by a factor $\mu(x,y)$, the ODE becomes a
total derivative\footnote{In this paper we use the term ``integrating
factor" in connection with the explicit form of the ODE, i.e., the ODE,
turned exact by taking the product $\mu \, ODE$, is assumed to be of the
form $\yt=\Phi(x,y,\y1)$ or $\yt-\Phi(x,y,\y1)=0$.},

\begin{equation}
\mu(x,y)\l(\y1-\Phi(x,y) \r) = \frac{d}{dx}\ R(x,y)
\label{exact_o1}
\end{equation}

\ni for some function $R$, then one can look for $\mu$ as a solution to
the first order PDE:

\begin{equation}
{\frac {\pa \mu}{\pa x}}
+{\frac {\pa }{\pa y}}\medl( \mu\,\Phi \medr)=0
\label{PDE_mu_1st_order}
\end{equation}

\ni which arises as the exactness condition for the problem (see \eq{EC}).
Once $\mu$ has been obtained, $R(x,y)$ - an implicit form solution - can be
calculated as a line integral.

Although to solve \eq{PDE_mu_1st_order} for $\mu$ is as difficult as the
original problem, it turns out that for a given $N(x,y)$, when a solution of
the form $\mu(x,y)=\ \mut(q)\,N(x,y)$ exists - $q$ is either $x$ or $y$ only
- $\mu$ can be determined by solving an auxiliary linear first order ODE.
For example, introducing $\mu(x,y)=\ \mut(x)\,N(x,y)$ and $M(x,y)=N(x,y)\,
\Phi(x,y)$, one obtains:

\begin{equation}
\mut(x)=C_1\,
{e^{^{\ \displaystyle
    -{\int \!
        {\frac {1}{N}}
        \l(
        {{\frac {\pa M}{\pa y}}}
        +{{\frac {\pa N}{\pa x}}}
        \r)
    {dx}}
}}}
\end{equation}

\ni and a solution $\mut(x)$ exists only when the integrand in above does
not depend on $y$. This gives both an existence condition and an explicit
solution to the problem; however, the advantages of the scheme are only
apparent since there is no way to determine in advance what would be the
appropriate $N(x,y)$.

\subsection{High order ODEs}

Integrating factors for high order ODEs are defined as in the first order
case. Here, we consider $\mu(x,y,\y1,...,\ym)$ to be an integrating factor
for an \nth order ODE, say

\begin{equation}
% {\frac {d^ny}{dx^n}}=\Phi(x,y,\y1,...,\ym)
\yn=\Phi(x,y,\y1,...,\ym)
\label{oh}
\end{equation}

\ni if after multiplying the explicit ODE by $\mu$ we obtain a total
derivative:

\begin{equation}
% \mu\l({\frac {d^ny}{dx^n}}-\Phi\r) = \frac{dR}{dx}
\mu\l(\yn-\Phi\r) = \frac{dR}{dx}
\label{exact_oh}
\end{equation}

\ni for some function $R(x,y,\y1,...,\ym)$. To determine $\mu$, one can try
to solve for it in the exactness condition, obtained applying Euler's
operator to the total derivative $H \equiv \mu\l(\yn-\Phi\r)$:

\begin{equation}
{\frac {\pa H}{\pa y}}-{\frac {d}{dx}}\l(\frac{\pa H}{\pa \y1}\r)
+{\frac {d^{2}}{d{x}^{2}}} \l( \frac{\pa H}{\pa \yt} \r)
+ ...
+ (-1)^n
{\frac {d^{n}}{d{x}^{n}}} \l( \frac{\pa H}{\pa y^{(n)}} \r) =0
\label{EC}
\end{equation}

\ni Now, it can be shown by induction that \eq{EC} is always of the form

\begin{equation}
A(x,y,\y1,...,y^{(2n-3)}) + y^{(2n-2)} B(x,y,\y1,...,\ym) = 0
\label{PDE_ABn}
\end{equation}

\ni where $A$ is of degree $n$-1 in $\yn$ and linear in $\yk$ for $n < k
\leq (2n-3)$, so that \eq{PDE_ABn} can be split into an overdetermined
system of PDEs for $\mu$. For example, for second order ODEs \eq{EC} is of
the form

\begin{equation}
A(x,y,\y1)+  \yt\ B(x,y,\y1)=0
\label{PDE_AB}
\end{equation}

\ni Hence, by taking $A(x,y,\y1)=0$ and $B(x,y,\y1)=0$ we have a system of
two PDEs for $\mu$:

\begin{eqnarray}
\lefteqn{A(x,y,\y1) \equiv}
& &
\label{PDE_A}
\\
& \mbox{\hspace{1cm}}&
\left ({\frac {\pa ^{2}\mu}{\pa \y1\pa x}}
+ \left ({\frac {\pa^{2}\mu}{\pa \y1\pa y}} \right )\y1
-{\frac {\pa \mu}{\pa y}}\right)\Phi
+\left ({\frac { \pa^{2}\Phi}{\pa \y1\pa x}}
-{\frac {\pa \Phi}{\pa y}}
+\left ({\frac { \pa ^{2}\Phi}{\pa \y1\pa y}}\right )\y1\right )\mu
\nonumber \\*[.1in]
& &
+\left({\frac {\pa ^{2}\mu}{ \pa {y}^{2}}}\right ){\y1}^{2}
+
\l(
\left ({ \frac {\pa \mu}{\pa \y1}}\right ){\frac {\pa \Phi} {\pa y}}
+\left({\frac {\pa \mu}{\pa y}}\right ){\frac { \pa \Phi}{\pa \y1}}
+2\,\left ({\frac {\pa ^{2}\mu}{ \pa x\pa y}}\right )
\r)
\y1
\nonumber \\*[.1in]
& &
+\left ({\frac {\pa \mu}{\pa \y1}} \right){\frac {\pa \Phi}{\pa x}}
+\left ({\frac {\pa \mu}{\pa x}}\right ){\frac {\pa \Phi}{\pa \y1}}
+{\frac {\pa^{2}\mu} {\pa {x}^{2}}}
= 0
\nonumber
\end{eqnarray}

\begin{eqnarray}
\lefteqn{B(x,y,\y1) \equiv }
& &
\label{PDE_B}
\\
&  \mbox{\hspace{1cm}} &
2\,{\frac {\pa \mu}{\pa y}}
+\left ({\frac { \pa ^{2}\mu}{\pa {\y1}^{2}}}\right )\Phi
+2 \,\left ({\frac {\pa \mu}{\pa \y1}}\right){\frac {\pa \Phi}{\pa \y1}}
+\mu{\frac {\pa ^{ 2}\Phi}{\pa {\y1}^{2}}}
+{\frac {\pa ^{2}\mu}{\pa \y1\pa x}}
+\left ({\frac {\pa^{2}\mu}{\pa \y1 \pa y}}\right )\y1
= 0
\nonumber
\end{eqnarray}

\ni Nonetheless, there are no general rules which might help in solving
these PDEs\footnote{In a recent work by \cite{bluman_paper} (1997),
the authors arrive at \eq{PDE_AB} and \eq{PDE_B} - numbered there as (3.5)
and (3.8) - departing from the adjoint linearized system corresponding to a
given ODE; the possible splitting of \eq{PDE_ABn} into an overdetermined
system for $\mu$ is also mentioned. However, in formula (3.5) of that work,
$\yt$ of \eq{PDE_AB} above appears replaced by $\Phi(x,y,\y1)$, and the
authors discuss possible alternatives to tackle Eqs.(\ref{PDE_AB}) and
(\ref{PDE_B}) instead of Eqs.(\ref{PDE_A}) and (\ref{PDE_B}).}.

Alternatively, a possible strategy for directly obtaining $R$ instead of
looking for $\mu$ can be formulated as follows. Consider the first order
linear operator associated to \eq{oh}

\begin{equation}
{\rm A :\ } f \rightarrow \frac{\pa f}{\pa x}
+ \y1 \frac{\pa f }{\pa y}
+ \yt \frac{\pa f }{\pa \y1}
+ ...
+ \Phi
\frac{\pa f}{\pa \ym}
\label{A}
\end{equation}

\ni where $x$, $y$ and its derivatives are all treated as independent
variables on the same footing. Now

\begin{equation}
A(R)=0
\label{A_R}
\end{equation}

\ni and there are $n$ functionally independent solutions (first integrals)
to the problem. In some cases, a first integral $R$ such that $dR/d\ym \neq
0$ can be obtained as the solution to a subset of the characteristic strip
of $A(R)=0$, or by other means.

\section{Integrating Factors and ODE patterns}
\label{intfactor}

Since the classical way for determining integrating factors leads to
problems similar in difficulty to solving the ODEs themselves, we consider
here a different approach, based on a careful matching of an ODE pattern.

The starting point is the observation that it is trivial to solve the
inverse problem; i.e., to find the most general ODE having a given $\mu$.
In fact, from \eq{exact_oh}, we have

\begin{equation}
\mu(x,y,\y1,...,\ym)=\fr{\pa\,R\mbox{\hspace{2mm}}}{\pa \ym}
\end{equation}

\ni and hence the reduced ODE $R$ is of the form

\begin{equation}
R=G(x,y,...,\ymm)+\int \mu\ d\ym
\label{R}
\end{equation}

\ni for some function $G$. Inserting \eq{R} into \eq{A_R} and solving for
$\yn$ leads to the general form of an ODE having $\mu$ as integrating
factor:

\begin{equation}
\label{ODE_mu}
\yn = \frac {-1}{\mu}
\l[{
    \frac {\pa }{\pa x}\l(\int \mu \, {d{\it \ym}} + G\r)
    + ...
    + \ym \frac {\pa }{\pa  \ymm}\l(\int \mu \, d\ym + G\r)
}
\r]
\end{equation}

\ni The expression above then becomes an {\it ODE pattern} which one can try
to match against an input ODE. The generation of such {\it pattern matching
routines} is difficult, even for restricted subfamilies of integrating
factor, but once built, they are a powerful and computationally efficient
way to reduce the order of the corresponding ODEs (see \se{tests}).

\subsection{Second order ODEs and the integrating factor family
$\mu(x,\y1)$}
\label{theorem}

In the case of second order ODEs, if instead of considering the general case
$\mu(x,y,\y1)$ we restrict the family of integrating factors under
consideration to $\mu(x,\y1)$, \eq{R} - the reduced ODE - becomes

\begin{equation}
\label{reduced1}
R(x,y,\y1)=F(x,\y1)+G(x,y)
\end{equation}

\ni for some functions $G$ and $F$, where

\begin{equation}
\label{mu_F}
{\mu}(x,\y1)= F_{\y1}(x,\y1)
\end{equation}

\ni (we denote $F_{\y1}=\frac{\pa F}{\pa \y1}$). \eq{ODE_mu} can then be
written in terms of $F$ and $G$ as

\begin{equation}
\label{reducible1}
\yt = \Phi(x,y,\y1) \equiv
- \frac{F_x(x,\y1) + G_x(x,y) + G_y(x,y)\, \y1}{F_{\y1}(x,\y1)}
\end{equation}

\ni The idea is now to build a routine to determine if a given ODE can be
written in the form \eq{reducible1}, in which case it will have an
integrating factor of the form $\mu(x,\y1)$, and if so, determine it,
leading to the reduced ODE \eq{reduced1} by means of standard methods (see
for instance \cite{murphy} p.221). The feasibility of such a
computational routine is based on the following theorem.

\begin{theorem}

Given a nonlinear second order ODE

\begin{equation}
\label{ODE2}
\yt = \Phi(x,y,\y1)
\end{equation}

\ni $(\frac{\pa \Phi}{\pa y} \neq 0)$ \footnote{ODEs {\it missing y} may
also have integrating factors of the form $\mu(x,\y1)$, which cannot be
determined using the scheme here presented. However, such integrating
factors are not really relevant since these ODEs can always be reduced to
first order by a simple change of variables.} for which an integrating
factor of the form $\mu(x,\y1)$ exists, such an integrating factor can be
systematically determined without solving any differential equations.

\end{theorem}

\ni PROOF.
We divide the proof in two steps. In the first step we {\it assume} that,
given \eq{ODE2}, it is always possible to determine $\mu(x,\y1)$ up to a
factor depending on $x$; that is, to find some $\F(x,\y1)$ satisfying

\begin{equation}
{\F}(x,\y1)= \frac {\mu(x,\y1)}{\mut(x)}
\label{cal_F}
\end{equation}

\ni for some unknown function ${\mut(x)}$. We then prove that the knowledge
of ${\F}(x,\y1)$ is enough to determine $\mut(x)$ by means of a simple
integral, hence leading to the desired $\mu(x,\y1)$.

In a second step, we prove our assumption, that is, we show how to find
${\F}(x,\y1)$ satisfying \eq{cal_F}, concluding the proof of the theorem.

\subsubsection{Determination of $\mut(x)$ when $\F(x,\y1)$ is known}

Starting with the first aforementioned step, we assume that we can determine
$\F(x,\y1)$. It follows from Eqs.(\ref{mu_F}), (\ref{reducible1})
and (\ref{cal_F}) that

\begin{equation}
\frac {\pa}{\pa y}
\medl(\Phi(x,y,\y1)\ \F(x,\y1)\medr) =
\frac {G_{y\,x}(x,y) + G_{y\,y}(x,y)\, \y1}{\mut(x)}
\label{G_eq}
\end{equation}

\ni so that by taking coefficients of $\y1$ in $\frac {\pa \Phi}{\pa y} \F$
we obtain

\begin{eqnarray}
\varphi_1 & \equiv &
\Phi_y(x,y,\y1)\, \F(x,\y1) -\y1\, \frac{\pa}{\pa \y1}
\medl( \Phi_y(x,y,\y1)\, \F(x,\y1) \medr)
= \frac {G_{y\,x}(x,y)}{\mut(x)}
\nonumber\\*[.07 in]
\varphi_2 & \equiv &
\frac{\pa}{\pa \y1}
\medl( \Phi_y(x,y,\y1)\, \F(x,\y1) \medr)
= \frac {G_{y\,y}(x,y)}{\mut(x)}
\label{varphi_1_2}
\end{eqnarray}

\ni Similarly, we obtain

\begin{eqnarray}
\varphi_3 & \equiv &
-\frac {\pa}{\pa \y1} \medl( \Phi(x,y,\y1)\ \F(x,\y1) \medr)
=
\frac {F_{\y1\,x}(x,\y1) + G_{y}(x,y)}{\mut(x)}
\nonumber\\*[.07 in]
\varphi_4 & \equiv & \frac {\pa}{\pa \y1} \F(x,\y1)
= \frac {F_{\y1\,\y1}(x,\y1)}{\mut(x)}
\label{varphi_3_4}
\end{eqnarray}

Now, since \eq{ODE2} is nonlinear by hypothesis, either $\varphi_2$ or
$\varphi_4$ is different from zero, so that at least one of the pairs of
ratios $\{\varphi_1,\ \varphi_2\}$ or $\{\varphi_3,\ \varphi_4\}$ can be
used to determine $\mut(x)$ as the solution of an auxiliary first order
linear ODE. For example, if $\varphi_2 \neq 0$,

\begin{equation}
\frac {\pa}{\pa y} \l(\varphi_1(x,y)\ {\mut(x)} \r)
= \frac {\pa}{\pa x} \l(\varphi_2(x,y)\ {\mut(x)} \r)
\end{equation}

\ni and we obtain

\begin{equation}
\mut(x)  =
{e^{^{\ \displaystyle
    {\int \!
        {\frac {1}{\varphi_2}}
        \l(
            {{\frac {\pa \varphi_1}{\pa y}}}
            -{{\frac {\pa \varphi_2}{\pa x}}}
        \r)
    {dx}}
}}}
\label{mut_2}
\end{equation}

\ni If $\varphi_2 = 0$ then $\varphi_4 \neq 0$ and we obtain

\begin{equation}
\mut(x)  =
{e^{^{\ \displaystyle
    {\int \!
        {\frac {1}{\varphi_4}}
        \l(
            {{\frac {\pa \varphi_3}{\pa \y1}}}
            -{{\frac {\pa \varphi_4}{\pa x}}}
        \r)
    {dx}}
}}}
\label{mut_3}
\end{equation}

\ni Eqs.(\ref{mut_2}) and (\ref{mut_3}) alternatively give both an explicit
solution to the problem and an existence condition, since a solution
$\mut(x)$ - and hence an integrating factor of the form $\mu(x,\y1)$ -
exists if the integrand in \eq{mut_2} or \eq{mut_3} only depends on $x$.
$\triangle$

\medskip
\ni {\it Example:} Kamke's ODE 37.
\medskip

\begin{equation}
\yt =  -2\,y\, \y1 - f(x)\left (\y1+{y}^{2}\right ) + g(x)
\label{k37}
\end{equation}

\ni This example is interesting\footnote{For ODE 6.37, Kamke shows a
reduction of order to a general Riccati ODE, based on the theory for ODEs
having solutions with no movable critical points - see
\cite{painleve}, and \cite{ince} p 331.} because it has no point
symmetries for arbitrary $f(x)$ and $g(x)$ (see \se{mu_and_X}). For this
ODE, $\F(x,\y1)$ was determined (see \se{lemma}) as:

\begin{equation}
\F(x,\y1) = 1
\end{equation}

\ni from which (\eq{G_eq})

\begin{equation}
\frac {G_{y\,x}(x,y) + G_{y\,y}(x,y)\, \y1}{\mut(x)} = -2 y\, f(x) -2\y1
\end{equation}

\ni and then as in \eq{varphi_1_2} we obtain

\begin{eqnarray}
\varphi_1 & = & -2 y\, f(x)
\nonumber\\*[.07 in]
\varphi_2 & = & -2
\end{eqnarray}

\ni Using this in \eq{mut_2}, we get

\begin{equation}
\mut(x) = {e^{^{\ \displaystyle {\int \! {f(x)}\, {dx}} }}}
\label{mu_k37}
\end{equation}

\ni and so, from \eq{cal_F}, since $\F(x,\y1)=1$, $\mu(x,\y1)=\,\mut(x)$.

\subsubsection{Determination of $\F(x,\y1)$}
\label{lemma}

We now prove our assumption, that is, we show how to obtain a function
$\F(x,\y1)$ satisfying \eq{cal_F} from the knowledge of $\Phi(x,y,\y1)$, and
without solving any differential equations.

Since we have already assumed that the given ODE has an integrating factor
of the form $\mu(x,\y1)$, then there exist some functions $F(x,\y1)$ and
$G(x,y)$ such that it is possible to rewrite $\Phi(x,y,\y1)$ - the
right-hand-side (RHS) of the given ODE - as in \eq{reducible1}. We then
start by considering the expression

\begin{equation}
\label{Upsilon}
\Upsilon \equiv
\frac {\pa \Phi}{\pa y} =
- \frac {G_{x\,y}(x,y) + G_{y\,y}(x,y)\ \y1}{F_{\y1}(x,\y1)}
\end{equation}

\ni and the possible cases.

\smallskip
\ni {\underline{Case A}}
\smallskip

The first case happens when the ratio $G_{x\,y}(x,y) / G_{y\,y}(x,y)$ {\it
depends} on $y$; i.e., $G_{x\,y}(x,y)$ and $G_{y\,y}(x,y)$ are linearly
independent w.r.t $y$. To determine whether this is the case, note that we
cannot just analyze the mentioned ratio itself since it is unknown. However,
we can always select the factors of $\Upsilon$ containing $y$, and check if
this expression {\it also} contains \y1. If so, we just determine
$F_{\y1}(x,\y1)$ up to a factor depending on $x$, that is, the required
$\F(x,\y1)$, as the reciprocal of the factors of $\Upsilon$ which depend on
\y1 but not $y$. $\triangle$

%%%%%%%%%%%%%%%%%%%%%%%%%

\medskip
\ni {\it Example:} Kamke's ODE 226
\medskip

\ni This ODE is presented in Kamke's book already in exact form, so we start
by rewriting it in explicit form as

\begin{equation}
\yt={\frac {x^2 y\y1+x y^2}{\y1}}  % this was ode 226
% y\yt-{\y1}^{2}+\left (x+x{y}^{2}\right )\y1-y\left (1-{y}^{2}\right ) = 0
% above is ode[123]
\label{k226}
\end{equation}

\ni We determine $\Upsilon$ (\eq{Upsilon}) as

\begin{equation}
\Upsilon = \frac {x (x \y1+2 y)}{\y1}
\end{equation}

\ni The only factor of $\Upsilon$ containing $y$ is:

\begin{equation}
x \y1+2 y
\end{equation}

\ni and since this also depends on \y1, $\F(x,\y1)$ is immediately given by

\begin{equation}
\F(x,\y1) = \y1
\end{equation}

%%%%%%%%%%%%%%%%%%%%%%%%%

\smallskip
\ni {\underline{Case B}}
\smallskip

When the expression formed by all the factors of $\Upsilon$ containing $y$
does not contain \y1, it is impossible to determine {\it a priori} whether
one of the functions $\{G_{x\,y}(x,y),\ G_{y\,y}(x,y)\}$ is zero, or
alternatively their ratio does not depend on $y$. We then proceed by
assuming the former, build an expression for $\F(x,\y1)$ as in Case A, and
determine $\mut(x)$ as explained in the previous subsection. If this doesn't
lead to the required integrating factor, we then proceed as follows.

\smallskip
\ni {\underline{Case C}}
\smallskip

In this case, we assume that neither $G_{x\,y}(x,y)$ nor $G_{y\,y}(x,y)$
are zero and their ratio is a function of just $x$, so that we have

\begin{eqnarray}
G_{x\,y}(x,y) & = &v_1(x)\ w(x,y)
\nonumber\\*[.07 in]
G_{y\,y}(x,y) & = &v_2(x)\ w(x,y)
\label{rho_v}
\end{eqnarray}

\ni for some unknown functions $v_1(x)$ and $v_2(x)$, such that
\eq{Upsilon} can be factored as

\begin{equation}
\Upsilon = w(x,y)\ \frac{\l(v_1(x) + v_2(x)\ \y1\r)}{F_{\y1}(x,\y1)}
\label{case_C}
\end{equation}

%($v_{\y1} \neq 0$)

\ni for some function $w(x,y)$ which {\it can always be determined} as the
factors of $\Upsilon$ depending on $y$. To determine $F_{\y1}(x,\y1)$ up to
a factor depending on $x$, we then need to determine the ratio
$v_1(x)/v_2(x)$. For this purpose, from \eq{rho_v} we build
an auxiliary PDE for $G_y(x,y)$,

\begin{equation}
\label{case_C_pde}
G_{x\,y}(x,y) = \frac {v_1(x)}{v_2(x)}\, G_{y,y}(x,y)
\end{equation}

\ni The general solution of \eq{case_C_pde} is given by

\begin{equation}
\label{calG}
G_y(x,y) = \G\l(y + p(x)\r)
\end{equation}

\ni where $\G$ is an arbitrary function of its argument and for
convenience we introduced

\begin{equation}
p'(x)\equiv v_1(x)/v_2(x)
\end{equation}

\ni We can now determine $p'(x)$, that is, the ratio $v_1/v_2$ we were
looking for, as follows. Taking into account \eq{rho_v}, we arrive at

\begin{equation}
\label{case_1_G_2}
v_2(x)\, w(x,y) = \G'(y+p(x))
\end{equation}

\ni By taking the ratio between this expression and its derivative w.r.t
$y$ we obtain

\begin{equation}
\H(y+p(x))\equiv
\frac {\pa w}{\pa y} / w
= \frac {\G''(y + p(x))}{\G'(y + p(x))}
\label{cal_H}
\end{equation}

\ni that is, a function of $y+p(x)$ only, which we can determine since we
know $w(x,y)$. If $\H' \neq 0$, we obtain $p'(x)$ as

\begin{equation}
\label{p_prime}
p'(x)  =  \frac {\pa \H}{\pa x} / \frac {\pa \H}{\pa y}
=
\frac {\left (\frac {\pa ^2 w} {\pa y^2 \pa x}\right )\ w
-\left (\frac {\pa w}{\pa y}\right )
\frac {\pa w}{\pa x}}
{\left (\frac {\pa ^2 w}{\pa y^2}\right )\ w
-\left (\frac {\pa w}{\pa y}\right )^2}
\end{equation}

\ni Once we determined $p'(x)$, from \eq{case_C} we determine $\F(x,\y1)$ as

\begin{equation}
\F(x,\y1)= \frac{(p'+\y1)\ w}{\Upsilon}
\label{F_p_w_Upsilon}
\end{equation}

\ni where $\Upsilon$, $w(x,y)$ and $p'(x)$ are now all known. $\triangle$

%%%%%%%%%%%%%%%%%%%%%%%%%

\medskip
\ni {\it Example:} Kamke's ODE 136.
\medskip

We begin by writing the ODE in explicit form as

\begin{equation}
\yt = \frac {h(\y1)}{x-y}
\label{k136}
\end{equation}

\ni This example is interesting since the standard search for point
symmetries is frustrated from the very beginning: the determining PDE for
the problem will not split due to the presence of an arbitrary function of
$\y1$. Here $\Upsilon$ (\eq{Upsilon}) is determined as

\begin{equation}
\Upsilon = -\frac {h({\it \y1})}{(x-y)^2}
\end{equation}

\ni and $w(x,y)$ as

\begin{equation}
w(x,y) = \frac {1}{(x-y)^2}
\end{equation}

\ni Then $\H(y + p(x))$ (\eq{cal_H}) becomes

\begin{equation}
\H = \frac {2}{x-y}
\end{equation}

\ni and hence, from \eq{p_prime},  $p'(x)$ is

\begin{equation}
p'(x) = -1
\end{equation}

\ni so from \eq{F_p_w_Upsilon}:

\begin{equation}
\F(x,\y1)= \frac{1-\y1} {h({\it \y1})}
\end{equation}

%%%%%%%%%%%%%%%%%%%%%%%%%

\smallskip
\ni \underline{Case D}
\smallskip

We now discuss how to obtain $p'(x)$ when $\H'(y+p(x))=0$. We consider at
first the case in which $\H=0$, hence $\G''=0$, and so, recalling
\eq{calG}, we see that

\begin{equation}
G(x,y) = B_1\ (y + p(x))^2 + B_2\ (y + p(x)) + g(x)
\end{equation}

\ni for some function $g(x)$ and some constants $B_1$, $B_2$. Recalling
\eq{reducible1}, $\Phi(x,y,\y1)$ takes the form

\begin{equation}
\label{case_1_subcase}
\Phi(x,y,\y1) = - \frac {F_x(x,\y1) + g'(x) + (2 B_1\ (y + p(x)) + B_2)
(\y1+p'(x))}{F_{\y1}(x,\y1)}
\end{equation}

\ni We can now obtain explicit equations where the only unknown is $p(x)$ as
follows. First, from the knowledge of $\Upsilon$ and $\Phi$ we build the two
explicit expressions:

\begin{equation}
\label{Lambda1}
\Lambda \equiv \frac {1}{\Upsilon} = - \frac {F_{\y1}} {2 B_1\ (\y1+p'(x))}
\end{equation}

\ni and

\begin{equation}
\label{Psi1}
\Psi \equiv \frac {\Phi(x,y,\y1)}{\Upsilon} - y =
\frac {F_x + g'(x)} {2 B_1\ (\y1+p'(x))} + p(x) + \frac{B_2}{2 B_1}
\end{equation}

\ni It is now clear from \eq{Lambda1} and \eq{Psi1} that $\Lambda$ and
$\Psi$ are related by the following equation:

\begin{equation}
\label{case_1_subcase_part2}
\frac {\pa}{\pa x} \medl((\y1+p'(x))\ \Lambda\medr) +
\frac {\pa}{\pa \y1} \medl((\y1+p'(x))\ \Psi\medr)
= p(x) + \frac {B_2}{2 B_1}
\end{equation}

\ni where the only unknowns are $p(x)$, $B_1$, and $B_2$. By differentiating
the equation above w.r.t \y1 and $x$ we obtain two equations where the only
unknown is $p'(x)$:

% \ni  Differentiating
% with respect to $\y1$ removes the unknown constants $B_1$ and $B_2$:

\begin{eqnarray}
& \Lambda_{\y1} p''(x) + (\Lambda_{x \y1} + \Psi_{\y1 \y1}) (\y1+p'(x)) +
\Lambda_x + 2 \Psi_{\y1} = 0 &
\label{case_1_subcase_part3}
\\*[.11 in]
& \Lambda\ p'''(x) + (\Lambda_{x x} + \Psi_{\y1 x}) (\y1 + p'(x))
+ (\Lambda_x + \Psi_{\y1}) p''(x) + \Psi_x = p'(x) &
\label{case_1_subcase_part4}
\end{eqnarray}

\ni As a shortcut, if $(\Lambda_{x \y1} + \Psi_{\y1 \y1}) / \Lambda_{\y1}$
depends on \y1, then we can build a linear algebraic equation for $p'(x)$ by
solving for $p''(x)$ in \eq{case_1_subcase_part3} and differentiating w.r.t.
\y1. Otherwise, in general we obtain $p'(x)$ by solving a linear algebraic
equation built by eliminating $p''(x)$ between \eq{case_1_subcase_part3} and
\eq{case_1_subcase_part4}\footnote{From \eq{Lambda1}, $\Lambda \neq 0$, so
that \eq{case_1_subcase_part4} always depends on $p'''(x)$, and solving
\eq{case_1_subcase_part3} for $p''(x)$ and substituting twice into
\eq{case_1_subcase_part4} will lead to the desired equation for $p'(x)$. If
\eq{case_1_subcase_part3} depends on $p'(x)$ but not on $p''(x)$, then
\eq{case_1_subcase_part3} itself is already a linear algebraic equation for
$p'(x)$.}.

If \eq{case_1_subcase_part3} depends neither on $p'(x)$ nor on $p''(x)$ this
scheme will not succeed. However, it is possible to prove that in that case
the original ODE is already linear, and easy to solve. To see this, we set
to zero the coefficients of $p'(x)$ and $p''(x)$ in
\eq{case_1_subcase_part3}, obtaining:

\begin{equation}
\label{case_1_subcase_part5}
\Lambda_{\y1} = \Lambda_{x \y1} + \Psi_{\y1 \y1} = \Lambda_x + 2
\Psi_{\y1}=0
\end{equation}

\ni from which $\Lambda$ is a function of $x$ only, and then

\begin{equation}
\label{case_1_subcase_part6}
\Psi_{\y1 \y1} = 0
\end{equation}

\ni If we now rewrite $F(x,\y1)$ as in

\begin{equation}
\label{case_1_subcase_part7}
F(x,\y1) = Z(x,\y1) -  g(x) - \Lambda (\y1 + p')^2\,B_1
\end{equation}

\ni and introduce this expression in \eq{Lambda1}, we obtain $Z_{\y1} = 0$;
similarly, using this result, \eq{Psi1}, \eq{case_1_subcase_part6} and
\eq{case_1_subcase_part7} we obtain $Z_x=0$. Hence, $Z$ {\it is a constant}.
Finally, taking into account that $Z$ is constant, \eq{case_1_subcase_part7}
and \eq{case_1_subcase}, we see that the ODE \eq{ODE2} which led us to this
case is just a non-homogeneous linear ODE of the form

\begin{equation}
\label{case_1_subcase_part9}
(y+p)''+ (\Lambda' (y+p)' - 2 (y + p) - B_2/B_1) / 2 \Lambda=0
\end{equation}

\ni which homogeneous part does not depend on $p(x)$:

\begin{equation}
\yt+\frac{\Lambda'(x)}{2{\Lambda(x)}}\,\y1-{\frac {y}{\Lambda(x)}}=0
\end{equation}

\ni and which solution is in any case straightforward. $\triangle$

%%%%%%%%%%%%%%%%%%%%%%%%%

\medskip
\ni {\it Example:} Kamke's ODE 66.
\medskip

This ODE is given by

\begin{equation}
\yt = a\left (c+bx+y\right )\left ({\y1}^{2}+1\right )^{3/2}
\end{equation}

\ni Proceeding as in Case A, we determine $\Upsilon$, $w(x,y)$, and
$\H(y+p(x))$ as

\begin{equation}
\Upsilon = a\left (\y1^{2}+1\right )^{3/2};\ \ \ \ \ \ \
w(x,y) = 1;\ \ \ \ \ \ \
\H = 0
\end{equation}

\ni From the last equation we realize that we are in Case D. We determine
$\Lambda$ and $\Psi$ (Eqs. (\ref{Lambda1}), (\ref{Psi1})) as:

\begin{eqnarray}
\Lambda & = & \frac{1}{\left ({\y1}^{2}+1\right )^{3/2}\, a}
\nonumber\\*[.07 in]
\Psi & = & c + b\, x
\end{eqnarray}

\ni We then build \eq{case_1_subcase_part2} for this ODE:

\begin{equation}
\frac{p''(x)}{\left ({\y1}^{2}+1\right )^{3/2}\, a} +
c + b\, x = p(x) + \frac {B_2}{2 B_1}
\end{equation}

\ni Differentiating w.r.t. \y1 leads to \eq{case_1_subcase_part3}:

\begin{equation}
-3 \frac{p''(x)\, \y1}{\left ({\y1}^{2}+1\right )^{5/2}\, a} = 0
\end{equation}

\ni from which it follows that $p''(x) = 0$. Using this in
\eq{case_1_subcase_part4} we obtain:

% \begin{equation}
% b + \frac{p'''(x)}{\left ({\y1}^{2}+1\right )^{3/2}\, a} - p'(x) = 0
% \end{equation}
%
% \ni we obtain

\begin{equation}
p'(x) = b
\end{equation}

\ni after which \eq{F_p_w_Upsilon} becomes

\begin{equation}
\F(x,\y1)= \frac{\y1+b} {a \left ({\y1}^{2}+1\right )^{3/2}}
\end{equation}

%%%%%%%%%%%%%%%%%%%%%%%%%

\smallskip
\ni \underline{Case E}
\smallskip

We now show how to obtain $p'(x)$ when $\H'(y+p(x))=0$ and $\H = \G''/\G'$
is a constant, $B_1$, which is different from zero; so $\G'$ is an
exponential function of its argument $(y+p(x))$ and hence from \eq{calG}

\begin{equation}
G(x,y) = B_2 e^{(y+p(x))B_1} +  (y+p(x))B_3 + g(x)
\end{equation}

\ni for some constants $B_2$, $B_3$ and some function $g(x)$. In this
case, it is always possible to arrive at an algebraic equation for
$p'(x)$, though the case entails some subtleties. First of all,
$\Phi(x,y,\y1)$ will be of the form

\begin{equation}
\label{case_1_subcase_0}
\Phi(x,y,\y1) = - \frac {F_x(x,\y1) + g'(x)
+ \l(B_2 B_1 e^{ (y + p(x))B_1} + B_3\r)
\l(\y1+p'(x)\r)}{F_{\y1}(x,\y1)}
\end{equation}

\ni Now, taking advantage of the fact that we explicitly know $B_1$, we
build our first explicit expression by dividing $B_1 e^{y B_1}$ by
$\Upsilon$:

\begin{equation}
\label{Lambda2}
\Lambda \equiv - \frac {F_{\y1}} {B_2 e^{p(x)B_1}\ (\y1+p'(x))}
\end{equation}

\ni We now multiply $\Phi$ by $\Lambda$ and subtract $B_1 e^{B_1 y}$ to
obtain our second explicit expression:

\begin{equation}
\label{Psi2}
\Psi \equiv  \frac {1}{B_2 e^{p(x)B_1}} \l(\frac {F_x + g'(x)} {\y1+p'(x)}
+ B_3\r)
\end{equation}

\ni Now, as in Case D, $\Lambda$ and $\Psi$ are related by

\begin{eqnarray}
{\frac {\pa}{\pa x} \medl((\y1+p'(x))\,\Lambda\medr)
+ \l(\y1+p'(x)\r)\, p'(x) \Lambda B_1
+ \frac {\pa}{\pa \y1} \medl((\y1+p'(x))\,\Psi\medr)}
& &
\nonumber \\
=\frac {B_3}{B_2 e^{p(x)B_1}}
\label{case_1_subcase_0_part2}
\end{eqnarray}

\ni where the only unknowns are $B_2$, $B_3$ and $p(x)$. We build a
first equation for $p'(x)$ by differentiating \eq{case_1_subcase_0_part2}
with respect to $\y1$

\begin{eqnarray}
\lefteqn{
\label{case_1_subcase_0_part3}
\medl(p''(x) + {p'(x)}^2 B_1\medr) \Lambda_{\y1}
        + p'(x) \medl(\y1 \Lambda_{\y1} B_1 + \Lambda B_1
        + \Lambda_{x \y1} + \Psi_{\y1 \y1}\medr)
        }
\\
&\mbox{\hspace{5.5cm}} &
+ 2 \Psi_{\y1} + \Lambda_x + \y1 \Lambda_{x \y1}
+ \y1 \Psi_{\y1 \y1} = 0
\nonumber
\end{eqnarray}

\ni The problem now is that, due to the exponential on the RHS of
\eq{case_1_subcase_0_part2}, differently from Case D, we are not able to
obtain a second expression for $p'(x)$ by differentiating w.r.t $x$. The
alternative we have found to determine $p'(x)$ can be summarized as follows.

\ni We first note that if $\Lambda_{\y1}=0$, \eq{case_1_subcase_0_part3} is
already a linear algebraic equation for $p'$\footnote{We can see this by
assuming $\Lambda_{\y1}=0$ and that \eq{case_1_subcase_0_part3} does not
contain $p'$, and then arriving at a contradiction as follows. We first set
the coefficients of $p'$ in \eq{case_1_subcase_0_part3} to zero, arriving at

\begin{displaymath}
\label{case_1_subcase_0_part3_b}
\mbox{\hspace{3.5cm}}
0 = B_1\ \Lambda + \Psi_{\y1 \y1} = 2 \Psi_{\y1} + \Lambda_x
+ \Psi_{\y1 \y1}
\y1
\mbox{\hspace{3.5cm}}
{(A)}
\end{displaymath}

\ni Eliminating $\Psi_{\y1 \y1}$ gives
$$2 \Psi_{\y1} = B_1\ \Lambda\ \y1 - \Lambda_x$$

\ni Differentiating the expression above w.r.t \y1 and since
$\Lambda_{\y1}=0$ we have,

$$2 \Psi_{\y1 \y1} = B_1\ \Lambda$$

\ni Finally using Eq.(A), $0 = \Lambda$,
contradicting $F_{\y1} \neq 0$.}, so that we are only worried
with the case $\Lambda_{\y1} \neq 0$. With this in mind, we divide
\eq{case_1_subcase_0_part3} by $\Lambda_{\y1}$
%
%  \eq{case_1_subcase_0_part3} for
% $p''$
%
% \begin{equation}
% \label{case_1_subcase_0_part4}
% p''(x) + B_1\ {p'(x)}^2
%         + p'(x) \frac {B_1\ \Lambda_{\y1} \y1 + B_1\ \Lambda
%         + \Lambda_{x \y1} + \Psi_{\y1 \y1}}{\Lambda_{\y1}}
%         + \frac {2 \Psi_{\y1} + \Lambda_x + \Lambda_{x \y1} \y1
%         + \Psi_{\y1 \y1} \y1}{\Lambda_{\y1}} = 0
% \end{equation}
%
% \ni where the only unknown is $p'(x)$, and, {\it if} the expression above
and, {\it if} the resulting expression
depends on \y1, we directly obtain a linear algebraic equation in $p'(x)$
just differentiating w.r.t \y1.$\triangle$

%%%%%%%%%%%%%%%%%%%%%%%%%

\medskip \ni {\it Example:}

\medskip

\begin{equation}
\yt=\frac {\y1\,\left (x\y1+1\right )\left (-2+{e^y}\right )}{\y1 x^2+\y1-1}
\end{equation}

\ni This example is interesting because it involves a non-rational
dependency on $y(x)$ - the dependent variable - thus being out of the scope
of most of the symmetry analysis software presently available. It is also
curious that there are no examples of this type in all of Kamke's set of
non-linear second order ODEs. On the other hand, using the algorithm here
presented, proceeding as in Case A, we determine $\Upsilon$, $w(x,y)$, and
$\H(y+p(x))$ as

\begin{equation}
\Upsilon = \frac {\y1 (x \y1 + 1) e^y}{\y1 x^2 + \y1 - 1};\ \ \ \ \ \ \
w(x,y) = e^y;\ \ \ \ \ \ \
\H = 1
\end{equation}

\ni From the last equation we know that we are in Case E. We then determine
$\Lambda$ and $\Psi$ as in Eqs. (\ref{Lambda2}) and (\ref{Psi2}):

\begin{eqnarray}
\Lambda & = & \frac {\y1 x^2 + \y1 - 1}{\y1 (x \y1 + 1)}
\nonumber\\*[.07 in]
\Psi & = & -2
\end{eqnarray}

\ni Now, we build \eq{case_1_subcase_0_part2}:

\begin{equation}
\frac  {1}{x \y1+1}
\left(\left (p''+ {p'}^{2}+ {\y1}^2 \frac{(x p'-1)}{x \y1+1}\right)
\left (x^2 +1-\frac {1}{\y1} \right)+2\,x p'-2 \right)
=\frac {B_3}{B_2\, e^p}
\end{equation}

\ni and, differentiating w.r.t. \y1, we obtain
(\eq{case_1_subcase_0_part3}),

\begin{equation}
\frac {2\,x \y1+1-(x^3+x) {\y1}^2}{ {\y1}^2 (x \y1+1)^2}\left (p''+ {p'}^2
\right )+\frac {2\,\y1-1-2\,x+x \y1}{\left (x \y1+1\right )^3}(x p'-1)=0
\end{equation}

\ni Proceeding as explained, dividing by $\Lambda_{\y1}$ and differentiating
w.r.t. \y1 gives

\begin{equation}
\frac {\pa}{\pa \y1} \left({\y1}^2 \frac {2\,\y1-1-2\,x+x \y1}{\left (x
\y1+1\right ) (2\,x \y1+1-(x^3+x) {\y1}^2 )}\right)(x p'-1)=0
\end{equation}

\ni Solving for $p'(x)$ gives $p'(x)=1/x$, from which (\eq{F_p_w_Upsilon}):

\begin{equation}
\F(x,\y1)= \l(\y1-\frac{1}{x}\r)\frac {\y1 x^2 + \y1 - 1}{\y1 (x \y1 + 1)}
\end{equation}

%%%%%%%%%%%%%%%%%%%%%%%%%

\smallskip
\ni {\underline{Case F}}
\smallskip

The final branch occurs when \eq{case_1_subcase_0_part3} divided by
$\Lambda_{\y1}$ does not depend on $\y1$ (so that we will not be able to
differentiate w.r.t \y1). In this case we can build a linear algebraic
equation for $p'(x)$ as follows. Let us introduce the label
$\beta(x,p',p'')$ for \eq{case_1_subcase_0_part3} divided by
$\Lambda_{\y1}$, so that \eq{case_1_subcase_0_part3} becomes:

\begin{equation}
\label{Lambda_beta}
\Lambda_{\y1}(x,\y1)\ \beta(x,p',p'') = 0
\end{equation}

\ni Again, since we obtained \eq{case_1_subcase_0_part3} by differentiating
\eq{case_1_subcase_0_part2} with respect to \y1, we see that
\eq{case_1_subcase_0_part2} can be written in terms of $\beta$ by
integrating \eq{Lambda_beta} with respect to \y1:

\begin{equation}
\Lambda(x,\y1) \beta(x,p',p'') + \gamma(x,p',p'') = \frac {B_3} {B_2
e^{p(x) B_1}}
\label{Lambda_beta_gamma}
\end{equation}

\ni where $\gamma(x,p',p'')$ is the constant of integration, and can be
determined explicitly in terms of $x$, $p'$ and $p''$ by comparing
\eq{Lambda_beta_gamma} with \eq{case_1_subcase_0_part2}. Taking into account
that ${\beta}(x,p',p'')=0$, we see that \eq{Lambda_beta_gamma} reduces to:

\begin{equation}
\label{gamma}
\gamma(x,p',p'') = \frac{B_3}{B_2 e^{p(x)B_1}}
\end{equation}

\ni We can remove the unknowns $B_2$ and $B_3$ after multiplying \eq{gamma}
by $e^{p(x)B_1}$, differentiating with respect to $x$, and then dividing
once again by $e^{p(x)B_1}$. We now have our second equation for $p'$, which
we can build explicitly in terms of $p'$, since we know $\gamma(x,p',p'')$
and $B_1$:

\begin{equation}
\label{gamma_p}
\frac {d \gamma}{dx}+B_1\ p' \gamma=0
\end{equation}

\ni Eliminating the derivatives of $p'$ between \eq{Lambda_beta} and
\eq{gamma_p} leads to a linear algebraic equation in $p'$. Once we have
$p'$, the determination of $\F(x,\y1)$ follows directly from
\eq{F_p_w_Upsilon}.
$\Box$

\subsection{Integrating factors of the form $\mu(y,\y1)$}

Just as in the previous section, from \eq{ODE_mu}, the ODE family admitting
an integrating factor of the form $\mu(y,\y1)$ is given by

\begin{equation}
\label{reducible2}
\yt=-{\frac {\y1}{\mu(y,\y1)}}
\l(\int \!\mu_{{y}}{d\y1}+G_{{y}}\r)-{\frac {G_{{x}}}{\mu(y,\y1)}}
\end{equation}

% the following first integral is
% associated with the integrating factor family $\mu(y,\y1)$:
%
% \begin{equation}
% \label{reduced12}
% R(x,y,\y1)= F(y,\y1)+ G(x,y)
% \end{equation}
%
% \ni where $\mu = F_{\y1}$, and to this integrating factor corresponds the
% ODE pattern:
%
% \begin{equation}
% \label{reducible2}
% \yt = - \frac{ G_x(x,y) + (F_y(y,\y1) + G_y(x,y)) \y1} {F_{\y1}(y,\y1)}
% \end{equation}

\ni For this ODE family, it would be possible to build a pattern matching
routine as done in the previous section for the case $\mu(x,\y1)$. However,
it is straightforward to notice that under the transformation $y(x)
\rightarrow x,\ x \rightarrow y(x)$, \eq{reducible2} transforms into an ODE
of the form \eq{reducible1} with integrating factor $\mu(x,\y1^{-1})/\y1^2$.
This means that the above pattern can be matched by merely changing
variables in the given ODE and matching \eq{reducible1}. It follows that any
explicit 2nd order ODE having an integrating factor of the form $\mu(y,\y1)$
can be reduced to a first order ODE by first changing variables, and then
using the scheme outlined in the previous section (unless the resulting ODE
is linear).

\medskip
\ni {\it Example:}
\medskip

\begin{equation}
\yt -{\frac {{\y1}^{2}}{y}} + \sin(x)\,\y1\,y +\cos(x)\,{y}^{2} = 0
\label{gon}
\end{equation}

\ni Changing variables as in $y(x) \rightarrow x,\ x \rightarrow y(x)$ we
obtain

\begin{equation}
\yt+{\frac {\y1}{x}}-\sin(y)\,{\y1}^{2}x-\cos(y)\,{x}^{2}{\y1}^{3}=0
\label{gon_y_x}
\end{equation}

\ni Using the algorithm outlined in the previous section, an
integrating factor of the form $\mu(x,\y1)$ for \eq{gon_y_x} is given by

\begin{equation}
{\frac {1}{{\y1}^{2}x}}
\end{equation}

\ni from where an integrating factor of the form $\mu(y,\y1)$ for \eq{gon}
is $1/y$, leading to the first integral

\begin{equation}
\sin(x)y+{\frac {\y1}{y}}+C_1=0,
\end{equation}

\ni which is a first order ODE of Bernoulli type. The solution to \eq{gon}
then follows directly. This example is particularly interesting since from
\cite{gonzales} we know ODE \eq{gon} has no point symmetries.

\subsection{Integrating factors of the form $\mu(x,y)$}

For completeness, we review here the determination of integrating factors of
the form $\mu(x,y)$ for second order ODEs, already found in the literature
(see for instance Lemma 3.8 in \cite{sheftel}). Contrary to the
cases $\mu(x,\y1)$ or $\mu(y,\y1)$, an integrating factor depending only on
$x$ and $y$ can easily be found - when it exists - by directly solving the
determining equations (\ref{PDE_A}) and (\ref{PDE_B}).

>From \eq{ODE_mu}, the general second order ODE having an integrating factor
$\mu(x,y)$ takes the form

\begin{equation}
\label{mu_xy_ode}
\yt = a(x,y)\, \y1 ^2 + b(x,y)\, \y1 + c(x,y),
\end{equation}

\ni where

\begin{equation}
\label{mu_xy_ode_cond}
a(x,y) = - \frac{\mu_y}{\mu},\ \ \
b(x,y) = - \frac{G_y + \mu_x}{\mu},\ \ \
c(x,y) = - \frac{G_x}{\mu}
\end{equation}

\ni for some unknown function $G(x,y)$. As a shortcut to solving Eqs.
(\ref{PDE_A}) and (\ref{PDE_B}), one can directly tackle
Eqs.(\ref{mu_xy_ode_cond}); the calculations are straightforward. There are
two cases to be considered.

\medskip
\ni {\underline {Case A}}: $\ 2a_x-b_y \neq 0$
\medskip

\ni Defining the two auxiliary quantities

\begin{equation}
\varphi \equiv c_y - a\,c - b_x,\ \ \ \ \
\Upsilon \equiv a_{x,x} + a_x\,b + \varphi_y
\label{auxiliary}
\end{equation}

\ni an integrating factor of the form $\mu(x,y)$ exists only when

\begin{equation}
\Upsilon_y - a_x = 0,\ \ \ \ \ \
\Upsilon_x + \varphi + b\,\Upsilon - \Upsilon^2 = 0
\end{equation}

\ni and is then given by

\begin{equation}
\mu(x,y)=
\exp\l(
\int \l(-\Upsilon+ \fr{\pa}{\pa x} \int a\, dy \r) dx
- \int a\, dy
\r)
\end{equation}

\medskip
\ni {\underline {Case B}}: $\ 2a_x-b_y = 0$
\medskip

\ni Redefining $\varphi \equiv c_y - a\,c$, an integrating factor of the
form $\mu(x,y)$ exists only when

\begin{equation}
a_{x,x} - a_x\,b - \varphi_y = 0,
\end{equation}

\ni Then, $\mu(x,y)$ is given by

\begin{equation}
\mu(x,y)=\nu(x)\ {\rm e}^{{^{-\displaystyle \int  a\, dy}}}
\end{equation}

\ni where $\nu(x)$ is either one of the independent solutions of the second
order linear ODE

\begin{equation}
\nu'' = A(x)\, \nu' + B(x)\, \nu,
\label{lode}
\end{equation}

\ni and
\begin{equation}
{\cal I} \equiv \fr{\pa}{\pa x} \int a\, dy,\ \ \ \ \ \
A(x) \equiv 2\, {\cal I} - b,\ \ \ \ \ \
B(x) \equiv
\varphi
+ \l(
{\cal I} - \fr{\pa}{\pa x}
\r)
\l(
    b- {\cal I}
\r)
\end{equation}

\ni It should be noted that when the attempt to solve the linear ODE
\eq{lode} is successful, using each of its two independent solutions for
integrating factors leads to the general solution of \eq{mu_xy_ode}, instead
of just a reduction of order. Also, when the original ODE was linear,
\eq{lode} is just the corresponding adjoint equation, as was to be expected
(see for instance Murphy's book).

\subsection{The Connection to PDEs}

Let $R(x,y,\y1)$ be a first integral of \eq{ODE2}. We rewrite \eq{A_R} by
renaming $\y1\equiv z$

\begin{equation}
\label{solvable_PDE}
\frac {\pa  R} {\pa  x} + z \frac {\pa  R} {\pa  y}
+ \Phi(x,y,z) \frac {\pa  R} {\pa  z} = 0
\end{equation}

\ni From Theorem \ref{theorem}, if a given PDE of the form
\eq{solvable_PDE} has a particular solution of the form $R(x,y,z) = F(x,z)
+ G(x,y)$, such that $R(x,y,z)$ is nonlinear in $y$ or $z$; or $R(x,y,z) =
F(y,z) + G(x,y)$, such that $R(x,y,z)$ is nonlinear in $y$ or $z^{-1}$,
then $F$ and $G$ can be determined in a systematic manner.

Although this is a natural consequence of the previous sections, it is worth
mentioning that the determination of R using the scheme here presented {\it
does not require} solving the characteristic strip of \eq{solvable_PDE},
thus being a genuine alternative.

\section{Integrating factors and symmetries}
\label{mu_and_X}

The main result being presented in this paper is a systematic algorithm for
the determination of integrating factors of the form $\mu(x,\y1)$ and
$\mu(y,\y1)$ {\it without solving any auxiliary differential equations}, and
this last fact is the most relevant point. Nonetheless, it is interesting to
briefly review the similarities and differences between the standard
integrating factor ($\mu$) and symmetry approaches, so as to have an insight
of how complementary these methods can be in practice.

To start with, both methods tackle an {\it n$^{th}$} order ODE by looking
for solutions to a linear {\it n$^{th}$ order determining PDE} in $n+1$
variables (see \se{introduction}). Any given ODE has infinitely many
integrating factors and symmetries. When many solutions to these {\it
determining} PDEs are found, both approaches can, in principle, give a
multiple reduction of order.

In the case of integrating factors there is one unknown function, while for
symmetries there is a pair of infinitesimals to be found. But symmetries are
defined up to an arbitrary function, so that we can always take one of these
infinitesimals equal to zero; hence we are facing approaches of equivalent
levels of difficulty and actually of equivalent solving power too.

Also valid for both approaches is the fact that, unless some {\it
restrictions} are introduced on the functional dependence of $\mu$ or the
infinitesimals, there is no hope that the corresponding determining PDEs
will be easier to solve than the original ODE. In the case of symmetries, it
is usual to restrict the problem to ODEs having {\it point symmetries}, that
is, to consider infinitesimals depending only on $x$ and $y$. The
restriction to the integrating factors here discussed is similar: we
considered $\mu$'s depending on only two variables.

At this point it can be seen that the two approaches are complementary: the
determining PDEs for $\mu$ and for the symmetries are different\footnote{We
are considering here ODEs of order greater than one.}, so that even using
identical restrictions on the functional dependence of $\mu$ and the
infinitesimals, problems which may be untractable using one approach may be
easy or even trivial using the other one.

As an  example of this, consider Kamke's ODE 6.37, appearing in this
paper as \eq{k37}:

\begin{displaymath}
\yt + 2\,y\, \y1 + f(x)\left (\y1+{y}^{2}\right ) - g(x) = 0
\end{displaymath}

\ni As mentioned in the exposition, for arbitrary $f(x)$ and $g(x)$, this
ODE has an integrating factor depending only on $x$, easily determined using
the algorithm presented. Now, for non-constant $f(x)$ and $g(x)$, this ODE
has no point symmetries, that is, no solutions of the form $[\xi(x,y),\
\eta(x,y)]$, except for the particular case in which $g(x)$ can be expressed
in terms of $f(x)$ as in\footnote{To determine $g(x)$ in terms of $f(x)$ we
used the {\it standard form} Maple package by Reid and Wittkopf complemented
with some basic calculations.}

\begin{equation}
g(x)=
\fr{\it f''}{4}\,
+ \fr{3\,f\,{\it f'}}{8}
+ \fr{f^3}{16}
-{\frac {C_2\,{\exp{\l(-3/2\,\displaystyle\int \!f(x){dx}\r)}}}
{4\,\left (2\,C_1+ \displaystyle
\int \!{\exp\l({-1/2\,\int \!f(x){dx}}\r)}\,{dx}\right )^{3}}}
\end{equation}

\ni Furthermore, this ODE has no non-trivial symmetries of the form
$[\xi(x,\y1),\ \eta(x,\y1)]$ either, and for symmetries of the form
$[\xi(y,\y1),\ \eta(y,\y1)]$ the determining PDE does not even split into a
system.

Another ODE example of this type is found in a paper by
\cite{gonzales} (1988):

\begin{equation}
\yt - {\frac {{\y1}^{2}}{y}} - g(x)\,p\,{y}^{p}\y1 - {\it g'}\,{y}^{p+1} = 0
\label{gon2}
\end{equation}

\ni In that work it is shown that for constant $p$, the ODE above only has
point symmetries for very restricted forms of $g(x)$. For instance, \eq{gon}
is a particular case of the ODE above and has no point symmetries.
Nonetheless, {\it for arbitrary} $g(x)$, \eq{gon2} has an obvious
integrating factor depending on only one variable: $1/y$, leading to a first
integral of Bernoulli type:

\begin{equation}
{\frac {\y1}{y}}-g(x){y}^{p}+C_1=0
\end{equation}

\ni so that the whole family \eq{gon2} is integrable by quadratures.

We note that \eq{k37} and \eq{gon2} are respectively particular cases of the
general reducible ODEs having integrating factors of the forms\footnote{To
obtain the general ODE family reducible by a given integrating factor we
used the routine {\bf redode} also presented in this paper in \se{commands}}
$\mu(x)$:

\begin{equation}
\yt=-{\frac {\left (\mu_{{x}}+G_{{y}}\right )}{\mu(x)}}\y1-{\frac {G_{
{x}}}{\mu(x)}}
\end{equation}

\ni where $\mu(x)$ and $G(x,y)$ are arbitrary; and $\mu(y)$:

\begin{equation}
\yt=-{\frac {( \mu_{{y}}\,\y1 + {G_{{y}}}) }{\mu(y)}} \,{\y1}
-{\frac {G_{{x}}}{\mu(y)}}
\end{equation}

\ni In turn, these are very simple cases if compared with the general ODE
families \eq{reducible1} and \eq{reducible2}, respectively having
integrating factors of the forms $\mu(x,\y1)$ and $\mu(y,\y1)$, and which
can be systematically reduced in order using the algorithm here presented.

It is then natural to conclude that the integrating factor and the symmetry
approaches can be useful for solving different types of ODEs, and can be
viewed as equivalently powerful and general, and in practice complementary.
Moreover, if for a given ODE, an integrating factor and a symmetry are
known, in principle one can combine this information to build two first
integrals and reduce the order by two at once (\cite{stephani}, chap.
3).

\section{Tests}
\label{tests}

After plugging the reducible-ODE scheme here presented into ODEtools, we
tested the scheme and routines using Kamke's non-linear 246 second order ODE
examples\footnote{Kamke's ODEs 6.247 to 6.249 cannot be made explicit and
are then excluded from the tests.}. The purpose was to confirm the
correctness of the returned results and to determine which of these ODEs
have integrating factors of the form $\mu(x,\y1)$ or $\mu(y,\y1)$. The test
consisted of determining $\mu$ and testing the exactness condition \eq{EC}
of the product $\mu$ times ODE.

We then ran a comparison of performances in solving a related subset of
Kamke's examples using different computer algebra ODE-solvers (Maple,
Mathematica, MuPAD and the Reduce package Convode). The idea was to situate
the new scheme in the framework of a sample of relevant packages presently
available. As a secondary goal, we were also interested in comparing the
solving performance of the new scheme with the one of the symmetry scheme
implemented in ODEtools.

Finally we considered the table of integrating factors for second order
non-linear ODEs found in Murphy's book and the answers for them returned by
all these ODE-solvers.

\subsection{The {\it reducible-ODE} solving scheme and Kamke's ODEs}

To run the test with Kamke's ODEs, the first step was to classify these ODEs
into: {\it missing x}, {\it missing y}, {\it exact} and {\it reducible},
where the latter refers to the new scheme. The reason for such a
classification is that ODEs missing variables are straightforwardly
reducible, so they are not the relevant target of the new scheme. Also, ODEs
already in exact form can be easily reduced after performing a simple check
for exactness; before running the tests all these ODEs were rewritten in
explicit form by isolating $\yt$. For classifying the ODEs we used the {\bf
odeadvisor} command from ODEtools. All the integrating factors found
satisfied the exactness condition \eq{EC}. The classification we obtained
for these 246 ODEs is as follows

{\begin{center} {\footnotesize
% \begin{tabular}{|p{1.14 in}|p{3.81 in}|}
\begin{tabular}{|p{1.14 in}|p{3.84 in}|}
\hline
Classification  &  ODE numbers as in Kamke's book \\
\hline
99 ODEs are missing $x$ or missing $y$ &
1, 2, 4, 7, 10, 12, 14, 17, 21, 22, 23, 24, 25, 26, 28, 30, 31, 32, 40,
42, 43, 45, 46, 47, 48, 49, 50, 54, 56, 60, 61, 62, 63, 64, 65, 67, 71,
72, 81, 89, 104, 107, 109, 110, 111, 113, 117, 118, 119, 120, 124, 125,
126, 127, 128, 130, 132, 137, 138, 140, 141, 143, 146, 150, 151, 153, 154,
155, 157, 158, 159, 160, 162, 163, 164, 165, 168, 188, 191, 192, 197, 200,
201, 202, 209, 210, 213, 214, 218, 220, 222, 223, 224, 232, 234, 236, 237,
243, 246 \\
\hline
13 are in exact form &
36, 42, 78, 107, 108, 109, 133, 169, 170, 178, 226, 231, 235
\\
\hline
% 68 ODEs are {\it reducible} with integrating factor
% $\mu(x,\y1)$ or $\mu(y,\y1)$ &
% 1, 2, 4, 7, 10, 12, 14, 17, 36, 37, 40, 42, 50, 51, 56, 64, 65, 66, 78,
% 81, 89, 97, 104, 107, 108, 109, 110, 111, 123, 125, 126, 133, 134, 135,
% 136, 137, 138, 150, 154, 155, 157, 164, 166, 168, 169, 173, 174, 175, 176,
% 178, 179, 188, 191, 192, 193, 196, 203, 204, 206, 209, 210, 214, 215, 218,
% 220, 222, 226, 235, 236 \\
%
40 ODEs are {\it reducible} with integrating factor
$\mu(x,\y1)$ or $\mu(y,\y1)$ {\it and}
missing $x$ or $y$&
1, 2, 4, 7, 10, 12, 14, 17, 40, 42, 50, 56, 64, 65, 81, 89, 104, 107, 109,
110, 111, 125, 126, 137, 138, 150, 154, 155, 157, 164, 168, 188, 191, 192,
209, 210, 214, 218, 220, 222, 236 \\
\hline
28 ODEs are {\it reducible} and not missing $x$ or $y$ &
36, 37, 51, 66, 78, 97, 108, 123, 133, 134, 135, 136, 166, 169, 173,
174, 175, 176, 178, 179, 193, 196, 203, 204, 206, 215, 226, 235
\\
\hline
\multicolumn{2}{c}{Table 1. Missing variables, exact and {\it
reducible} Kamke's 246 second order non-linear ODEs.}
\end{tabular}}
\end{center}}

>From the table above, $\approx 30 \%$ of these 246 ODEs from Kamke's book
are reducible to first order using the scheme here presented. Also, although
the symmetry scheme implemented in ODEtools - which works with dynamical
symmetries and includes heuristic procedures - finds symmetries for 191 of
these 246 ODEs, it is unsuccessful in reducing the order of five ODEs which
the new scheme does reduce. These are the ODEs numbered 36, 37, 123, 215,
and 235. It is interesting to note that ODEs 36, 37 and 123 have no point
symmetries; ODE 215 lead to a third order PDE system whose solution - in
terms of elliptic integrals - can be obtained by {\it computer plus hand} if
one uses trial and error; and for ODE 235, the determining PDE for the
symmetries does not split into a system due to the presence of an arbitrary
function of $\y1$. Also, none of the other computer algebra ODE-solvers
tested during this work succeeded in solving or reducing the order of any of
these five ODEs\footnote {A table of results obtained using the Reduce
package {\it CRACK}, kindly sent to us by Dr. T. Wolf, shows that out of
these five, {\it CRACK} is determining symmetries only for ODE 215, and
concluding that there exist no point symmetries for ODEs 6.36, 6.37, 6.123
and 6.235, but perhaps for some undetermined special values of the function
parameters entering the ODEs (see details for ODE 6.37 in \se{mu_and_X}).},
even though the corresponding integrating factors depend on only one
variable (see \se{comparison}).

For ODE 215, which we write in explicit form as

\begin{equation}
\ode
\yt
=
\frac{
\left (6\,y^{2}
-\frac{a}{2}\right )\y1^{2}}
{4\,y^{3}-a\,y-b}
- f(x)\,\y1
\end{equation}

\ni the integrating factor found by the computer algebra
implementation of the new scheme (see \se{commands}) is\footnote{In what
follows, the {\it input} can be recognized by the Maple prompt \verb->-.}:

\begin{verbatim}
> mu = intfactor(ode,y(x));
\end{verbatim}
\begin{equation}
\mu=\frac{1}{\y1}
\end{equation}

\ni This integrating factor leads to a reduced ODE which can be solved as
well, resulting in the following implicit solution in terms of an elliptic
integral\footnote{For ODE 6.215, there is a typographical mistake
in Kamke's book concerning the reduced ODE: instead of $...\sqrt{4y^3-g_2
y}-g_3...$, one should read $...\sqrt{4\,y^3-g_2 y-g_3}....$}:

\begin{verbatim}
> odsolve(ode);
\end{verbatim}
\begin{equation}
\int \!{e^{^{-\displaystyle\int \!f(x){dx}}}}{dx}-C_1\int ^{y}\!{\frac
{1}{\sqrt {-4\,{z}^{3}+z\,a+b}}}{dz}+C_2=0
\end{equation}

\ni The integrating factor found for ODE 37 (see Eqs.(\ref{k37}) and
(\ref{mu_k37})) leads to a reduction of order resulting in the most general
first order Riccati type ODE; in this example {\bf odsolve} just returns the
reduction of order obtained by the new integrating factor scheme. For ODE
123, the integrating factor found is $\displaystyle 1/y$, and the reduced
ODE is also a generic Riccati ODE. Finally, ODE 235 appears in Kamke's book
written in exact form, but the ODE is interesting because it contains three
arbitrary functions, of the first derivative, the dependent and the
independent variables, respectively. Such an arbitrary dependence makes this
ODE almost intractable for most computer algebra ODE-solvers and related
packages. We then first isolated the highest derivative as to make the ODE
{\it non-exact}

\begin{equation}
\ode
\yt= - \frac{\l(G\!\l(y\r)\y1+F(x)\r)}{H\!\!\l(\y1\r)}
\label{ODE_235}
\end{equation}

\ni The integrating factor here found is

\begin{verbatim}
> mu = intfactor(ode,y(x));
\end{verbatim}
\begin{equation}
\mu=H\!\!\l(\y1\r)
\end{equation}

Concerning timings, it is worth mentioning that in the specific subset of 28
Kamke's examples which are not missing variables, the average {\it time
consumed} by {\bf odsolve} in solving each ODE using the new scheme was
$2.5$ sec, while using symmetries this time jumps to $21$ sec. These tests
were performed using a Pentium 200, 64 Mb RAM, running Windows 95. In
summary: for these 28 ODEs having an integrating factor of the form
$\mu(x,\y1)$ or $\mu(y,\y1)$, the new scheme seems to be, on the average,
$\approx 10$ times faster than the symmetry scheme.

\subsection{Comparison of performances}
\label{comparison}

With the classification presented in Table 1. in hands, we used different
computer algebra systems to run a comparison of performances in solving
these ODEs having integrating factors of the form $\mu(x,\y1)$ or
$\mu(y,\y1)$. For our purposes, the interesting subset is the one comprised
of the 28 ODEs not already missing variables (see Table 1.). The results we
obtained are summarized in the following table\footnote{When building the
statistics for ODEtools, we passed to {\bf odsolve} the optional argument
\verb-[reducible]-, meaning: try the reducible scheme, and if it does not
solve the problem just give up. To solve the reduced ODE all of {\bf
odsolve}'s methods, including symmetries, were used. The input and output in
the respective format for all the packages tested are available in
http://dft.if.uerj.br/odetools/mu\_odes.zip.}:

{\begin{center}
{\footnotesize
\begin{tabular}
{p{.9cm}p{2.3cm}p{2.4cm}p{2.4cm}p{2.95cm}}
\hline
% & \multicolumn{5}{c} {Kamke's ODE numbers}
\multicolumn{5}{c} {Kamke's ODE numbers}
\\
\hline
    & Convode &  Mathematica 3.0 & MuPAD 1.3 &  ODEtools
\\
\hline
Solved: &
51, 166, 173, 174, 175, 176, 179.
&
78, 97, 108, 166, 169, 173, 174, 175, 176, 178, 179, 206.
&
78, 97, 108, 133, 166, 169, 173, 174, 175, 176, 179,
&
51, 78, 97, 108, 133, 134, 135, 136, 166, 169, 173, 174,
175, 176, 178, 179, 193, 196, 203, 204, 206, 215.
\\
\hline
{\it Totals:} & \centering 7  & \centering 12
& \centering 11 &
{\centering 22

\vspace{-1cm}
}
\\
\hline
Reduced: &  &  &  &
36, 37, 66, 123, 226, 235.
\\
\hline
{\it Totals:} & \centering 0 & \centering 0 & \centering 0 &
{\centering 6

\vspace{-1cm}
}
\\
\hline
\multicolumn{5}{c}
{Table 2. \it Performances in solving 28 Kamke's ODEs having an integrating
factor $\mu(x,\y1)$ or
$\mu(y,\y1)$}\\
\end{tabular}}
\end{center}}

As shown above, while the scheme here presented is finding first integrals
in all the 28 ODE examples, opening the way to solve 22 of them to the end,
the next scores are only 12 and 11 ODEs, respectively solved by Mathematica
3.0 and MuPAD 1.3.

Concerning the six reductions of order returned by {\bf odsolve}, it must
be said that neither MuPAD nor Mathematica provides a way to convey them,
so that perhaps their ODE-solvers are obtaining first integrals for these
cases but the routines are giving up when they cannot solve the problem to
the end.

Maple R4 is not present in the table since it is not solving any of these 28
ODEs. This is understandable since in R4 the only methods implemented for
high order non-linear ODEs are those for ODEs which are missing variables.
This situation is being resolved in the upcoming Maple R5, where the
ODEtools routines are included in the Maple library, and the previous
ODE-solver has been replaced by {\bf odsolve}\footnote{However, the scheme
here presented was not ready when the development library was closed; the
{\it reducible} scheme implemented in Maple R5 is able to determine, when
they exist, integrating factors only of the form $\mu(\y1)$.}.

Although the primary goal of this work is just to obtain first integrals for
second order ODEs, it is also interesting to comment on the six ODEs shown
in Table 2.\/ for which the new scheme succeeds in determining integrating
factors but the reduced ODEs remain unsolved. First of all, for ODEs 36, 37
and 123 the reduction of order lead to general Riccati type ODEs, so that in
these cases no more than a reduction of order should be expected. Concerning
ODE 235 (\eq{ODE_235}), the reduced ODE is:

\begin{equation}
\int ^{\y1}\! H(z){dz}
+\int ^{y}\! G(z){dz} + \int \! F(x){dx}+C_1=0
\end{equation}

\ni Methods for solving such a first order ODE are known only for very
special explicit combinations of $H$, $G$ and $F$. Concerning ODEs 66 and
226, the obtained reduced ODEs are the same as those shown in Kamke's book,
and are out of the scope of {\bf odsolve}.

\subsection{The {\it reducible-ODE} scheme and Murphy's table of integrating
factors}

There is an explicit paragraph in Murphy's book concerning integrating
factors of the form $\mu(\y1)$, where it is shown a table with four second
order non-linear ODE families for which $\mu(\y1)$ is already known. The
first two families are trivial in the sense that they are already missing
variables. The third of these ODE families is:

\begin{equation}
\ode \yt=P(x)\,\y1+Q(y)\, \y1^{2}
\label{murphi1}
\end{equation}

\ni where $P$ and $Q$ are arbitrary functions of its arguments; this is
actually Liouville's ODE. The integrating factor mentioned in the book is
the same found by the scheme here presented: \y1; and the corresponding
reduced ODE can be solved in implicit form:

\begin{verbatim}
> odsolve(ode);
\end{verbatim}
\begin{equation}
\int \!{e^{\int \!P(x){dx}}}{dx}-\int ^{y}\!{e^{-\int \!Q(z
){dz}+C_1}}{dz}+C_2=0
\end{equation}

\ni The fourth ODE family is the most general second order ODE
having 1/\y1 as integrating factor (see \se{redode}):
\begin{equation}
\ode \yt=\frac{\pa R(x,y)}{\pa x}\y1+\frac{\pa R(x,y)}{\pa y}\ \y1^{2}
\label{murphi2}
\end{equation}

\ni for some function $R(x,y)$. Here the new scheme finds the integrating
factor $1/\y1$ and returns the reduced ODE

\begin{equation}
\ln (\y1)-R(x,y)+C_1=0
\end{equation}

\ni actually a generic first order ODE\footnote{For the third ODE family,
Mathematica 3.0 returns a wrong answer and MuPAD 1.3 gives up, while for the
fourth family, Mathematica gives up and MuPAD returns an ERROR message.}.

\section{Computer algebra implementation}
\label{commands}

We implemented the scheme for finding integrating factors described in
\se{intfactor} in the framework of the ODEtools package \cite{odetools2},
taking advantage of its set of programming tool routines specifically
designed to work with ODEs. The implementation consists of:

\begin{description}

\item[$\bullet$] The plugging of the {\it reducible-ODE} solving scheme here
presented in the block of methods for nonlinear second order ODEs of the
ODEtools command {\bf odsolve};

\item[$\bullet$] The extension of the capabilities of the ODEtools {\bf
intfactor} command to determine integrating factors for non-linear second
order ODEs using the scheme here presented;

\item[$\bullet$] A new user-level routine, {\bf redode}, returning the most
general explicit ODE having a given integrating factor (\eq{ODE_mu});

\end{description}

The computational implementation follows straightforwardly the explanations
of \se{intfactor} and includes three main routines, for determining
$\F(x,\y1)$, $\mut(x)$ and the reduced ODE $R(x,y,\y1)$, respectively.
Callings to these routines were in turn added to both the {\bf intfactor}
and {\bf odsolve} commands, so that the scheme becomes available at
user-level.

A test of this implementation in {\bf odsolve} and some related examples are
found in \se{tests}. Since detailed descriptions of the ODEtools commands
are found in the On-Line help, we have restricted this section to a
description of the new command {\bf redode} followed by two examples.

\subsection*{\it Description of {\bf redode}}
\label{description}

% The main purpose of the {\bf redode} command is to use it as an auxiliary
% tool in developing solving methods. This subsection summarizes this new
% routine.

\medskip
\ni {\it Command name: {\bf redode}}
\label{redode}

\ni {\it Feature:} returns the \nth order ODE having a given integrating
factor

\ni {\it Calling sequence:}

\ni \verb-> redode(mu, n, y(x))-;

\ni \verb-> redode(mu, n, y(x), R)-;

\medskip
\ni {\it Parameters:}

\noindent
\begin{tabular}{p{0.5in}p{4.35in}}

\verb-n- & - indicates the order of the requested ODE.\\

\verb-mu- & - an integrating factor depending on $x$, $y,...,\ym$.\\

\verb-y(x)- & - the dependent variable.\\

\verb-R- & - optional, the expected reduced ODE depending on $x$,
$y,...,\ym$.\\

\end{tabular}

\medskip
\noindent
{\it Synopsis:}
\smallskip

$\bullet$ Given an integrating factor $\mu(x,y,...\yn)$, {\bf redode}'s
main goal is to return the ODE of order $n$ having $\mu$ as integrating
factor

\begin{displaymath}
\yn = \frac {-1}{\mu}
\l[{
    \frac {\pa }{\pa x}\l(\int \mu \, {d{\it \ym}} + G\r)
    + ...
    + \ym \frac {\pa }{\pa  \ymm}\l(\int \mu \, d\ym + G\r)
}
\r]
\end{displaymath}

\ni where $G\equiv G(x,y,...\ymm)$ is an arbitrary function of its arguments
(see \se{intfactor}). This command is useful to identify the general ODE
problem related to a given $\mu$, as well as to understand the possible
links between the integrating factor scheme for reducing the order and other
reduction schemes (e.g., symmetries).

$\bullet$ When the expected {\it reduced ODE} (differential order $n-$1),
here called $R$, is also given as argument, the routine proceeds as follows.
First, a test to see if the requested ODE exists is performed:

\begin{equation}
{\mu(x,y,...\ym)}=\nu(x,y,...\ymm)\
{{\frac {\pa }{\pa \ym}}R(x,y,...,\ym)}
\end{equation}

\ni for some function $\nu(x,y,...\ym)$. If the problem is solvable, {\bf
redode} will then return an \nth order $ODE^{(n)}= \yn - \Phi(x,y,...,\ym)$
satisfying

\begin{equation}
\mu(x,y,...,\ym)\ ODE^{(n)}=\frac{d}{dx}
\bigl(\nu(x,y,...\ymm)\ R(x,y,...,\ym) \bigr)
\end{equation}

\ni that is, an ODE having as first integral $\nu\, R+ \mbox{\it constant}$.

% \ni Note that the reduced ODE being differentiated on the right-hand-side
% is defined up to a constant - i.e., $R+C_1$ satisfies the equation above
% just as well.

$\bullet$ When the given $\mu$ does not depend on $\ym$ and $R$ is
non-linear in $\ym$, the requested \nth order ODE nevertheless exists if $R$
can be solved for $\ym$.

% - then $\nu$ can be determined as the ratio $\mu /
% \frac{\pa R}{\pa \ym}$.

\medskip
\ni {\it Examples:}
\medskip

The {\bf redode} command is interesting mainly as a tool for generating
solving schemes for given ODE families; we illustrate with two examples.

\smallskip 1. Consider the family of second order ODEs having as integrating
factor $\mu=F(x)$ - an arbitrary function - such that the reduced ODE has
the same integrating factor. We want to set up an algorithm such that, given
a second order linear ODE,

\begin{equation}
\yt = \psi_1(x)\, \y1 + \psi_2(x)\, y + \psi_3(x)
\label{ode2}
\end{equation}

\ni where there are no restrictions on $\psi_1(x),\ \psi_2(x)$ or
$\psi_3(x)$, the scheme determines if the ODE belongs to the family just
described, and if so it also determines $F(x)$. The knowledge of $F(x)$ will
be enough to build a closed form solution for the ODE.

To start with we obtain the first order ODE having $F(x)$ as integrating
factor via

\begin{verbatim}
> ode_1 := redode(F(x), y(x));
\end{verbatim}
\begin{equation}
{\rm ode_1 := }\
\y1=-
{\frac{1}{F(x)}}
\l(y\ {\frac {dF(x)}{dx}}+{\it \_F1}(x)\r)
\end{equation}

\ni where ${\it \_F1}(x)$ is an arbitrary function. To obtain the second
order ODE aforementioned we pass \verb-ode_1- as argument (playing the role
of the {\it reduced} ODE) together with the integrating factor $F(x)$ to
obtain

\begin{verbatim}
> ode_2 := redode(F(x), y(x), ode_1);
\end{verbatim}
\begin{equation}
{\rm ode_2 := }\
\yt
=-{\frac{1}{F(x)}\l(
2\,\y1\
{\frac {dF(x)}{dx}}
+y{\frac {d^{2}F(x)}{d{x}^{2}}}
+{\frac {d{\it \_F1}(x)}{dx}}\r)}
\label{mu_Fx}
\end{equation}

% \ni The answer to \verb-ode_2- is obtained from the knowledge of its
% integrating factor $F(x)$ as
%
% \begin{equation}
% y={\frac {C_1\,x+C_2}{F(x)}}
% \end{equation}

Taking this general ODE pattern as departure point, we setup the required
solving scheme by comparing coefficients in \eq{ode2} and \eq{mu_Fx},
obtaining

\begin{equation}
{\frac {-2}{F(x)}{{\frac {dF(x)}{dx}}}}={\psi_1}(x),\ \ \ \ \ \ \
{\frac {-1}{F(x)}{{\frac {d^{2}F(x)}{d{x}^{2}}}}}=\psi_2(x)
\label{e1}
\end{equation}

\ni By solving the first equation, we get $F(x)$ as

\begin{equation}
F(x)=C_1\,{e^{^{\displaystyle-\int \!\frac{\psi_1(x)}{2}{dx}}}}
\end{equation}

\ni and by substituting this result into the second one we get the pattern
identifying the ODE family
\begin{equation}
\frac{1}{2}\,{\frac
{d\psi_1(x)}{dx}}-\frac{1}{4}\,\l({\psi_1}(x)\r)^{2}-{\psi_2}(x)=0
\label{e2}
\end{equation}

\ni Among the ODE-solvers of Maple R4, Mathematica 3.0, MuPAD 1.3 or
Convode (Reduce), only those of MuPAD and Maple succeed in solving this
ODE family.

\medskip 2. Consider the second order ODE family having as integrating
factor $\mu=F(x)$ - an arbitrary function - {\it also having} the
symmetry\footnote{Here we denote the infinitesimal symmetry generator by
$[\xi,\eta]\equiv \xi\frac{\pa}{\pa x} + \eta \frac{\pa}{\pa y}$}
$[\xi=0,\eta=F(x)]$, and such that the reduced ODE is the most general first
order linear ODE

\begin{equation}
{\rm ode_1 := }\
\y1= A(x)\,y+B(x)
\end{equation}

\ni where $A(x)$ and $B(x)$ are arbitrary functions. To start with, we
obtain the aforementioned second order ODE having the integrating factor
$F(x)$ as in Example 1.

\begin{verbatim}
> ode_2 := redode(F(x), y(x), ode_1);
\end{verbatim}
\begin{equation}
%{\rm ode_2 := }
\yt
=
\left ({\frac {dA(x)}{dx}}+{\frac {
\left ({\frac {dF(x)}{dx}}\right )A(x)}{F(x)}}\right )y
+\left (A(x)-{\frac {{\frac {dF(x)}{dx}}}{F(x)}}\right )\,
\y1
+{\frac {dB(x)}{dx}}+{\frac {\left ({\frac {dF(x)}{dx}}\right )B(x)}
{F(x)}}
\label{mu_Fx_2}
\end{equation}

\ni In this step, \verb-ode_2- is in fact the most general second order
linear ODE. If we now impose the symmetry condition\footnote{For linear
ODEs, symmetries of the form $[0,F(x)]$ are also symmetries of the
homogeneous part.} $X$(\verb-ode_2-)=0, where $X=[0,\, F(x)]$ we arrive at
the following restriction on $A(x)$

\begin{equation}
-F(x){\frac {dA(x)}{dx}}-2\,
\left ({\frac {dF(x)}{dx}}\right )A(x)
+{\frac {\left ({\frac {dF(x)}{dx}}\right )^{2}}{F(x)}}
+{\frac {d^{2}F(x)}{d{x}^{2}}}=0
\end{equation}

\ni Solving this ODE for $A(x)$, introducing the result into \eq{mu_Fx_2}
and disregarding the non-homogeneous term (irrelevant in the solving
scheme) we obtain the homogeneous ODE family pattern:

\begin{equation}
\yt=
\l(
\frac{1}{2}\,{\frac {dH(x)}{dx}}
+ \frac{3}{4}\,{\frac {\left ({\frac {dH(x)}{dx}}\right )^{2}}{\left
(H(x)\right )^{2}}}
-\frac{1}{2}\,{\frac {{\frac {d^{2}H(x)}{d{x}^{2}}}}{H(x)}}
\r)y
+H(x)\ \y1
\label{linear_2}
\end{equation}

\ni where we introduced $H(x)=(F(x))^{-2}$. Although this ODE family
appears more general than the one treated in Example 1., the setting up of
a solving scheme here is easier: one just needs to check if the
coefficient of $y$ in a given ODE is related to the coefficient of $\y1$
as in equation \eq{linear_2}, in which case the integrating factor is just
${{\frac {1}{\sqrt {H(x)}}}}$.

\section{Conclusions}
\label{conclusions}

This paper presented a systematic method for obtaining integrating factors
of the form $\mu(x,\y1)$ and $\mu(y,\y1)$ - when they exist - for second
order non-linear ODEs, as well as its computer algebra implementation in the
framework of the ODEtools package. The scheme is new, as far as we know, and
the implementation has proven to be a valuable tool since it leads to
reductions of order for varied ODE examples, as shown in \se{tests}.
Actually, the implementation of the scheme solves ODEs not solved by using
standard or symmetry methods (see \se{mu_and_X}) or other computer algebra
ODE-solvers (see \se{comparison}); furthermore, it involves only algebraic
operations, so that - in principle - it gives answers remarkably faster than
the symmetry scheme.

It is also worth mentioning that restricting the dependence of $\mu$ in
\eq{EC} to $\mu(x,\y1)$, {\it does not lead} to a straightforwardly solvable
problem except for few or simple cases. Moreover, when the determination of
$\mu$ from \eq{EC} is frustrated, there is no way to determine whether such
a solution $\mu(x,\y1)$ exists. It is then a pleasant surprise to see such
integrating factors - provided they exist - being systematically determined
in all cases and without solving any differential equations, convincing us
of the value of the new scheme.

On the other hand, we are restricting the problem to the universe of second
order ODEs having integrating factors depending only on two variables - the
general case is $\mu(x,y,\y1)$ - and even so, for integrating factors of the
form $\mu(x,y)$ the method may fail in solving the auxiliary linear ODE
\eq{lode} which appears in one of the subcases.

% Also, a routine, {\bf redode}, was designed to tackle the inverse problem,
% that is to find the most general \nth order ODE having a given integrating
% factor, optionally reducing to a given $k^{th}$ order ODE ($k<n$). This
% command can be useful in many situations, and for investigating new
% solving methods.

Some natural extensions of this work then would be to develop a scheme for
building integrating factors of the forms considered in this work, now for
higher order ODEs, at least for restricted ODE families yet to be
determined. Concerning these extensions, the {\bf redode} routine presented,
designed to find the most general \nth order ODE having a given integrating
factor, optionally reducing to a given $k^{th}$ order ODE ($k<n$), can be of
use in investigating further problems. We are presently working on these
possible extensions\footnote{See http://dft.if.uerj.br/odetools.html}, and
expect to succeed in obtaining reportable results in the near future.

\section*{Acknowledgments}

\noindent This work was supported by the State University of Rio de Janeiro
(UERJ), Brazil, and by the Symbolic Computation Group, Faculty of
Mathematics, University of Waterloo, Ontario, Canada. The authors would like
to thank K. von B\"ulow\footnote{Symbolic Computation Group, Theoretical
Physics Department, IF-UERJ-Brazil.} for a careful reading of this paper.

\newpage
\appendix
\section*{Appendix A}
% \hspace\parindent

This appendix contains some additional information which may be useful as a
reference for developing computer algebra implementations of this work, or
for improving the one here presented.

As explained in \se{lemma}, the scheme presented can be subdivided into six
different cases: A, B, C, D, E and F. Actually, there are just five cases
since case B is always either A or C. From the point of view of a computer
implementation of the scheme it is interesting to know what one would expect
from such an implementation concerning Kamke's ODEs and the cases
aforementioned. We then display here both the integrating factors obtained
for the 28 Kamke's ODEs used in the tests (see \se{tests}) and the ``case"
corresponding to each ODE.

{\begin{center} {\footnotesize
\begin{tabular}{|p{1 in}|p{3 in}|p{.35 in}|}
\hline
Integrating factor  &  Kamke's book ODE-number  & Case
\\
\hline
$1 $ &  36  &  D
\\
\hline
${e^{\int \!f(x){dx}}} $ &  37  &  A
\\
\hline
${\y1}^{-1} $ &  51, 166, 169, 173, 175, 176, 179, 196, 203, 204, 206,
                                                               215  &  C
\\
\hline
${\frac {b+\y1}{\left (1+{\y1}^{2}\right )^{3/2}}} $ &  66  &  D
\\
\hline
$x $ &  78  &  D
\\
\hline
${x}^{-1} $ &  97  &  A
\\
\hline
$y $ &  108  &  D
\\
\hline
${y}^{-1} $ &  123  &  A
\\
\hline
${\frac {1+\y1}{\left (\y1-1\right )\y1}} $ &  133  &  C
\\
\hline
${\frac {\y1-1}{\left (1+\y1\right )\y1}} $ &  134  &  C
\\
\hline
${\frac {\y1-1}{\left (1+\y1\right )\left (1+{\y1}^{2}\right )}} $
                                                             &  135  &  C
\\
\hline
${\frac {\y1-1}{h(\y1)}} $ &  136  &  C
\\
\hline
${\frac {x}{2\,x\y1-1}} $ &  174  &  C
\\
\hline
$\left (1+\y1\right )^{-1} $ &  178  &  C
\\
\hline
${\frac {1}{\y1\,\left (1+2\,y\y1\right )}} $ &  193  &  C
\\
\hline
$\y1 $ &  226  &  A
\\
\hline
$h(\y1) $ &  235  &  C
\\
\hline
\multicolumn{2}{c}
{Table A.1 Integrating factors for Kamke's {\it reducible} and
ODEs not missing variables.}
\end{tabular}}
\end{center}}
\label{lastpage}
\end{document}